\documentclass{article}

\usepackage{arxiv}

\usepackage[utf8]{inputenc} 
\usepackage[T1]{fontenc}    
\usepackage{hyperref}       
\usepackage{url}            
\usepackage{booktabs}       
\usepackage{amsfonts}       
\usepackage{nicefrac}       
\usepackage{microtype}      
\usepackage{lipsum}
\usepackage{graphicx}
\usepackage{upgreek}
\usepackage{stfloats}
\usepackage{amsmath}

\title{Optical design of a multi-object fiber-fed spectrograph system for Southern Spectroscopic Survey Telescope}

\author{
   YueFan Shan, ZhengBo Zhu, Hao Tan, Donglin Ma\thanks{Shenzhen Huazhong University of Science and Technology, Shenzhen 518057, China}   \\
  School of Optical and Electronic Information and Wuhan National Laboratory of Opto-electronics\\
  Huazhong University of Science and Technology\\
  Wuhan 430074, China \\
  \texttt{madonglin@hust.edu.cn} \\
  
}
\begin{document}
\maketitle
\begin{abstract}
Southern Spectroscopic Survey Telescope (SSST) is a wide-field spectroscopic survey telescope that China plans to build in Chile in the next few years. As an instrument for astronomical spectroscopic survey, the multi-object and fiber-fed spectrograph (MOFFS) is one of the most important scientific instruments for SSST. In this paper, we present a recommended optical design for the MOFFS system based on the Volume Phase Holographic Gratings (VPHG). The whole design philosophy and procedure, including the analytic method to determine the initial structure, optimization procedures of the VPHG and the camera groups, are demonstrated in detail. The numerical results of the final obtained spectrograph show a superior imaging quality and a relatively high transmittance for the whole working waveband and the field of view. The design method proposed in this paper can provide a reference for the design of MOFFS accommodated in spectroscopic survey telescopes.
\end{abstract}

\maketitle

\section{Introduction}
Image and spectroscopy are two of the most important observational information in astronomical research. For the former one,  a lot of telescopes have been designed for wide-area deep image sky survey observation, such as the Hubble Space Telescope (HST) \cite{1991The}, Large Synoptic Survey Telescope(LSST) \cite{2002Survey}, Chinese Space Station Telescope (CSST, under construction) \cite{2014Two} and James Webb Space Telescope (JWST, to be launched) \cite{2003Next}. As an indispensable complement, the southern multi-target spectral survey telescope will be the most important and widely used ground-based equipment for the next generation of wide-area image sky survey projects \cite{2016Maximizing}. For the spectroscopy survey, the huge success of Two-Degree Field Galaxy Redshift Survey (2dFGRS) \cite{Spergel2008First} and Sloan Digital Sky Survey (SDSS) \cite{Ivezic2002Optical} have proved that the large-field multi-object spectral survey telescope can maintain competitiveness and efficient scientific output for decades. However, the multi-object spectroscopic survey telescope with an aperture larger than 6 meters and field of view greater than 3 square degrees is a major gap of the parameter space for the current telescopes in service. Therefore, China has planed to build Southern Spectroscopic Survey Telescope (SSST) with an aperture of 6.5 meters and a field of view greater than $2^\circ\times2^\circ$ in Chile. SSST aims to perform an efficient spectral verification for wide-area image sky survey and reveal the physics behind astronomical events.

Two kinds of recommended optical designs for SSST have been proposed \cite{Tan:20,202065m}, which are designed for the telescope body. But the works about the spectrograph, which is also an important scientific instrument for SSST, have not been reported.

Up to now, the most powerful implemented design of such kind of instruments is the Dark Energy Spectroscopic Instrument (DESI), which was designed for the 3.8-meter Mayall telescope \cite{Aghamousa:2016sne}. It has a field of view of $3.2^\circ$  and can record the spectrum of 5000 different targets at the same time. The working waveband of DESI ranges from 360nm to 980nm. Each spectrograph of DESI consists of three channels and each channel is equipped with a 4k×4k CCD to obtain the high-resolution spectral information from 500 different objects. However, the statement of the design process was not detailed in \cite{Aghamousa:2016sne}.

In this paper, we aim to design a multi-object and fiber-fed spectrograph (MOFFS) system with a working waveband ranging from 360nm to 1100nm to cover both near-infrared and visible regions. To make the MOFFS work in a wider spectral range while maintaining a high resolution of spectral imaging, we divide the observing light into four sub-wavebands via 3 dichroic filters. For each sub-waveband, the specific optical channel is elaborately designed respectively, and each channel is equipped with a 4k×4k CCD with a pixel size of 15$\mu$m. The numerical result shows that the designed optical system has a superior optical performance for spectroscopy measurement.

\section{Principle}
\label{sec:examples}

\begin{figure}[ht!]
\includegraphics[width=16CM]{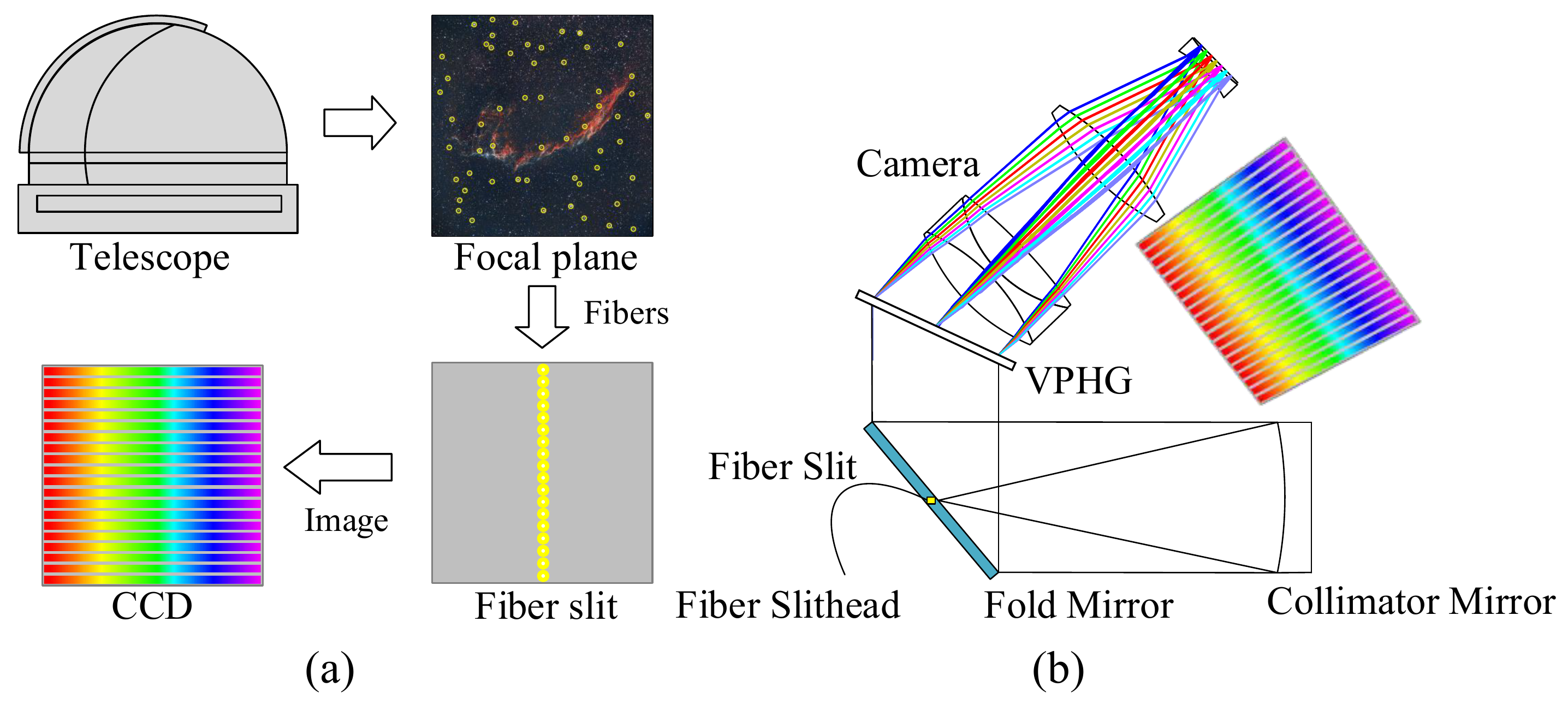}
\caption{(a) Schematic diagram of SSST; (b) optical layout of one channel of MOFFS.}
\label{fig:1}
\end{figure}

Figure \ref{fig:1} (a) illustrates the working mechanism of SSST, where light rays coming from the outer space are captured by the telescope firstly, and then these rays enter into one ends of fibers which are located at the focal plane of the telescope. The positions of these fiber's ends on the focal plane can be mechanically adjusted. The other ends of the fibers are mounted linearly forming a so-called fiber slit. Finally, the light rays emitted from the fiber slit are transmitted to the CCD via the spectroscopic imaging system.

 Figure \ref{fig:1}(b) shows the optical layout of one channel of the MOFFS system, which is composed of a fiber slit, a collimator mirror, a folder mirror, a dichroic filter (not drawn), a volume phase holographic grating (VPHG), and a camera unit. The light emitted from the fiber slit is firstly collimated by the collimator mirror. Then the light beam is divided into 4 channels corresponding to 4 sub-wavebands with the help of 3 dichroic filters. After that, the light is dispersed by the VPHG forming the mixed filed of views, and each field of view corresponds to a specific wavelength. Finally, the light beam 
propagates to CCD via the camera unit. Based on the relevant research experience and the specific application of SSST, we summarize the design technical specifications for the spectrometer optical system in Table \ref{tab:T1}.

\begin{table}[htbp]
\centering
\caption{\bf Design specification}
\begin{tabular}{ccc}
\hline
Wavelength range & 360nm - 1100nm \\
Spectral\ resolution & <0.06nm/pixel\\
Transmittance & >86\% (360nm - 390nm)\\
   & >90\% (390nm - 1100nm) \\
RMS\ radius & 15$\mu$m  \\
\hline
\end{tabular}
  \label{tab:T1}
\end{table}

\section{Determination of parameters}

\subsection{Determination of focal ratio}
Focal ratio degradation (FRD) is the most important factor to determine the throughput loss of the Fiber-fed spectrographs \cite{2017A,2014Studying}. Because of the FRD effect, the divergence angle of the beam emitted from the fiber will be larger than that of the beam entering into the fiber. The light rays beyond the acceptance solid angle of the spectrometer system lead to the FRD loss. So, the magnitude of the FRD effect is greatly determined by the focal ratio of the incident light \cite{2008Investigation,1988Focal}. According to Lawrence W. Ramsey's study \cite{1988Focal}, $F/\ 3\sim F/\ 4$ is the best choice of f-ratio to minimize the FRD effect. As a rule of thumb, we set the f-ratio as  $F/\ 3.6$ in our design.

\subsection{Determination of Fiber slit's parameters}
\begin{figure}[ht!]
\includegraphics[width=16CM]{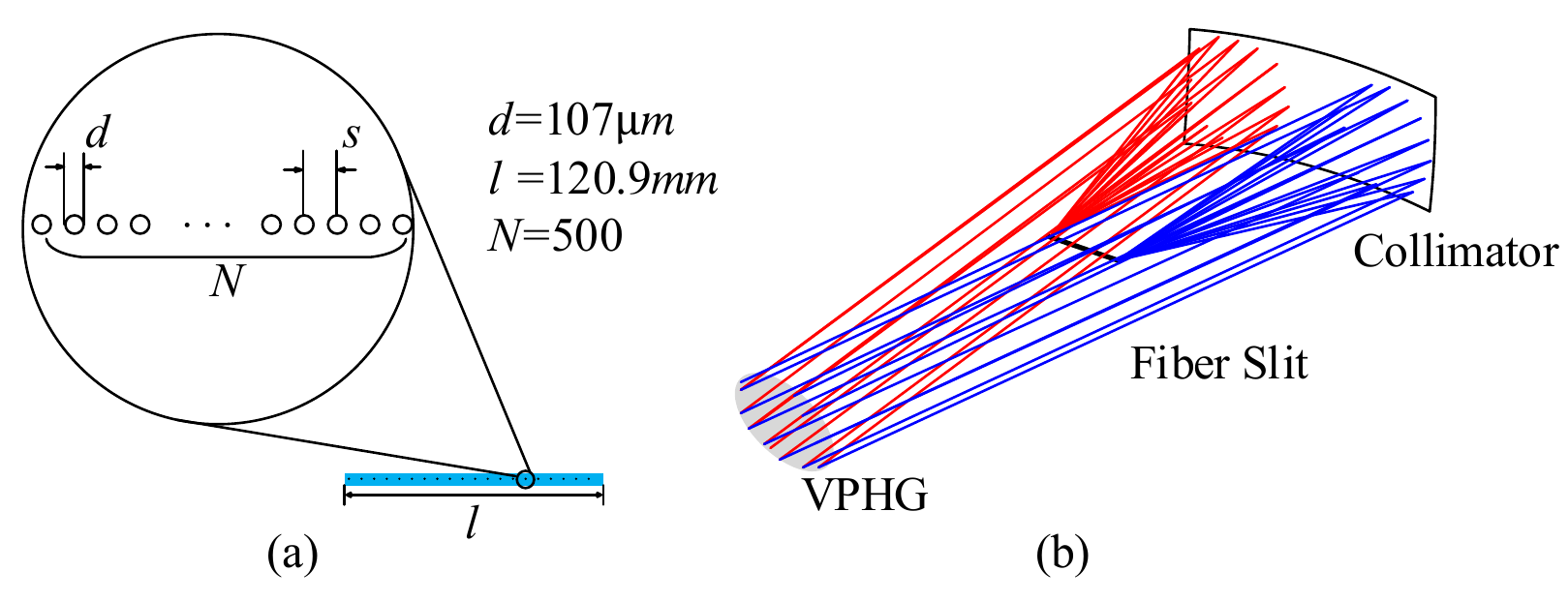}
    \caption{(a) Configuration of the fiber slit; (b) Schematic diagram of collimator system.}
\label{fig:2}
\end{figure}

The fiber slit is the object plane of MOFFS. The parameters of fiber slit are selected according to DESI \cite{Aghamousa:2016sne}. The slit length is 120.9mm and contains 500 fibers which are equally spaced, as shown in Fig. \ref{fig:2}(a).

\subsection{Determination of collimator's parameters}
\begin{figure}[ht!]
\includegraphics[width=6.5in]{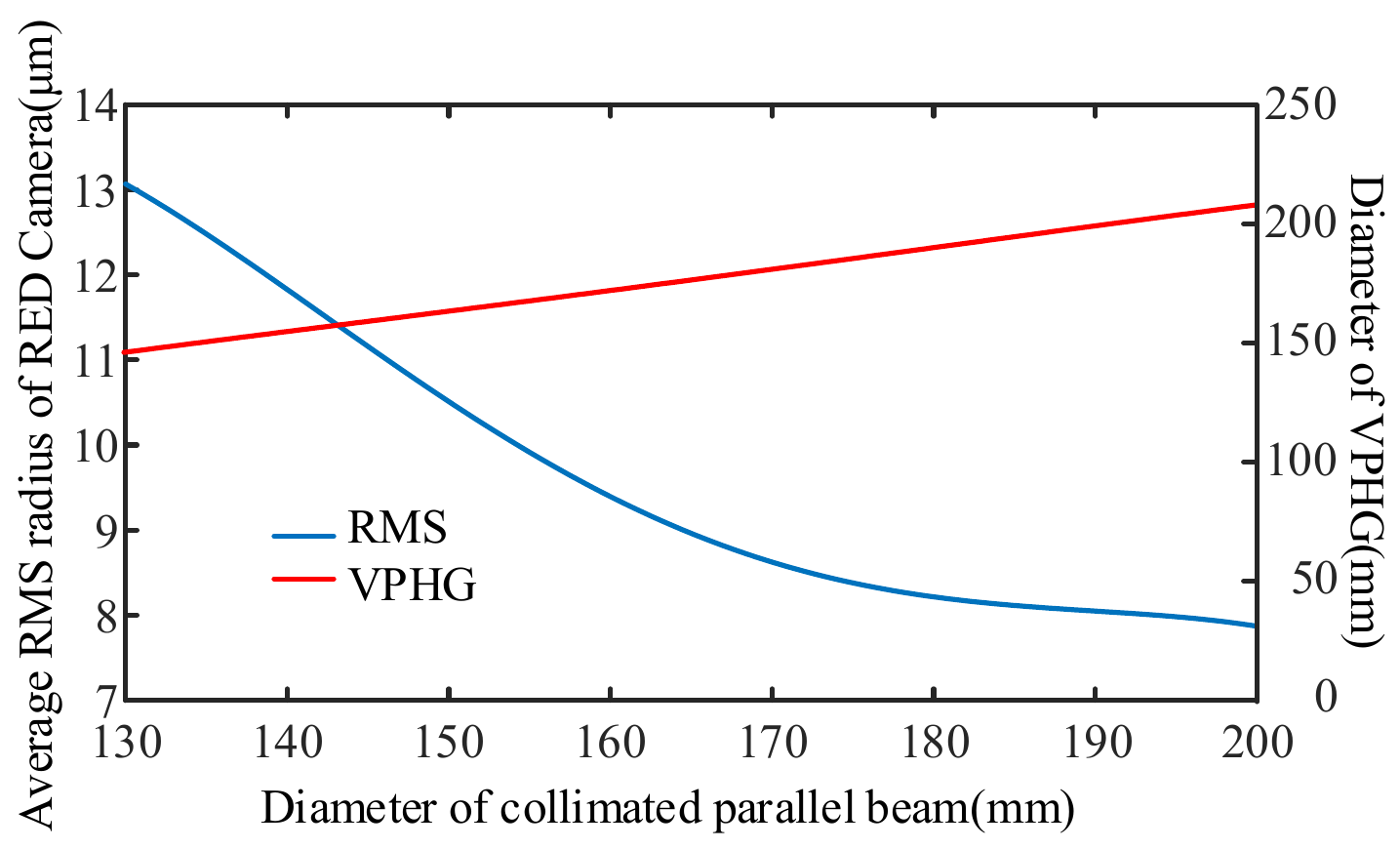}
\caption{Average RMS radius of RED Camera Vs diameter of collimated beam.}
\label{fig:3}
\end{figure}

In this design, a single spherical mirror is selected as the collimator as shown in Fig. \ref{fig:2}(b). The light rays emanating from both ends of the fiber slit are collimated to be a parallel light beam by the collimator, the angle between these two beams can be approximately calculated as

\[\tan \frac{\theta }{2} = \frac{l}{r}
\label{eq:1}
\tag{1},\]
where $\theta$ is the angle between the two beams, $l$ is the length of the fiber slit equaling 120.9mm, and $r$ is the radius of curvature of the collimator mirror. It is clear that the size of the VPHG is determined by the diameter of the collimated parallel beam. A larger collimated beam can definitely guarantee a better optical performance. While large collimated beam means high cost of VPHG. So, a trade-off between the optical performance and the cost of VPHG has to be made. As illustrated in Fig. \ref{fig:3}, a larger collimator makes a smaller average RMS radius, i.e. better optical performance. When the diameter of the collimator increases to 170mm, the descent rate of the average RMS radius tended to be an uneconomical level. So, we choose the diameter of collimator as 170mm in this design.

The radius of curvature of the collimator mirror can be obtained as follows
\begin{equation}
{F \mathord{\left/
 {\vphantom {F \# }} \right.
 \kern-\nulldelimiterspace} \# } \cdot D = \frac{r}{2}
\tag{2}
\label{eq:2}
\end{equation}
where $F/\#$ is the focal ratio of the beam emitted from the fiber slit, $D$ is the diameter of the collimator mirror. As mentioned above,  $F/\#$ is 3.6 and $D$  is 170mm. So, we can get $r$=1224mm and $\theta\approx11.28^\circ$. As a result, the distance from the fiber slit to the collimator mirror can be calculated as $D\cdot F/\# = 612$mm.

\subsection{Determination of dichroic's parameters}
\begin{figure}[ht!]
\includegraphics[width=6in]{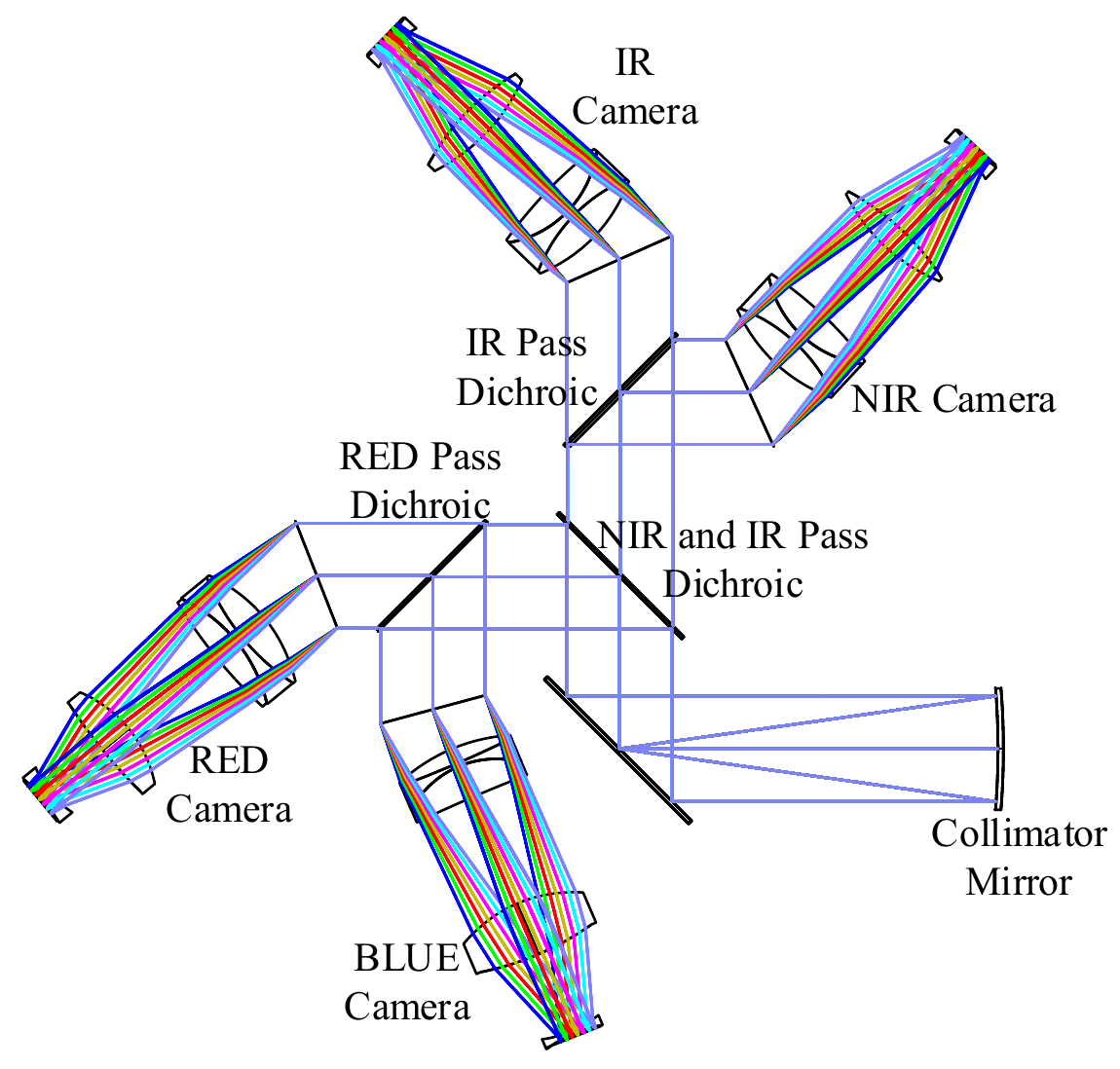}
\caption{Design layout of the whole system of multi-object fiber-fed spectrographs.}
\label{fig:4}
\end{figure}

For the waveband ranginig from $\lambda _1$ to $\lambda _2$, the average spectral resolution can be approximately expressed as
\begin{equation}
\Delta \lambda  = \frac{{\left( {{\lambda _2} - {\lambda _1}} \right)}}{N},
\tag{3}
\label{eq:3}
\end{equation}where $N$ is the number of pixels along the $x$ or $y$ direction of CCD. Considering the requirement of spectral resolution, we divide the whole waveband into four channels as listed in Table \ref{tab:T2}. The layout of the whole optical system is illustrated in Fig. \ref{fig:4}, and the corresponding parameters of the 3 dichroics are listed in Table \ref{tab:T3}.

\begin{table}[htbp]
\centering
\caption{\bf Waveband for each channel.}
\begin{tabular}{ccc}
\hline
Channel & Wavelength\ range\\
\hline
BLUE & 360nm-580nm\\
RED & 560nm-740nm\\
NIR & 720nm-900nm\\
IR & 877nm-1100nm\\
\hline
  \label{tab:T2}
\end{tabular}
\end{table}

\begin{table*}[htbp]
\centering
\caption{\bf Parameter requirements of dichroics.}
\begin{tabular}{cccc}
\hline
Dichroic & RED\ Pass& NIR\ and\ IR Pass& IR\ Pass\\
\hline
Transmission\ Band & 580nm-740nm & 740nm-1100nm & 900nm-1100nm\\
Reflection\ Band & 360nm -560nm & 360nm -720nm & 360nm -877nm\\
Crossover\ Region & 560nm -580nm & 720nm -740nm & 877nm -900nm\\
Crossover\ Width & 20nm & 20nm& 22.5nm\\
\hline
  \label{tab:T3}
\end{tabular}
\end{table*}

\subsection{Determination of VPHG's parameters}
\begin{figure}[ht!]
\includegraphics[width=6in]{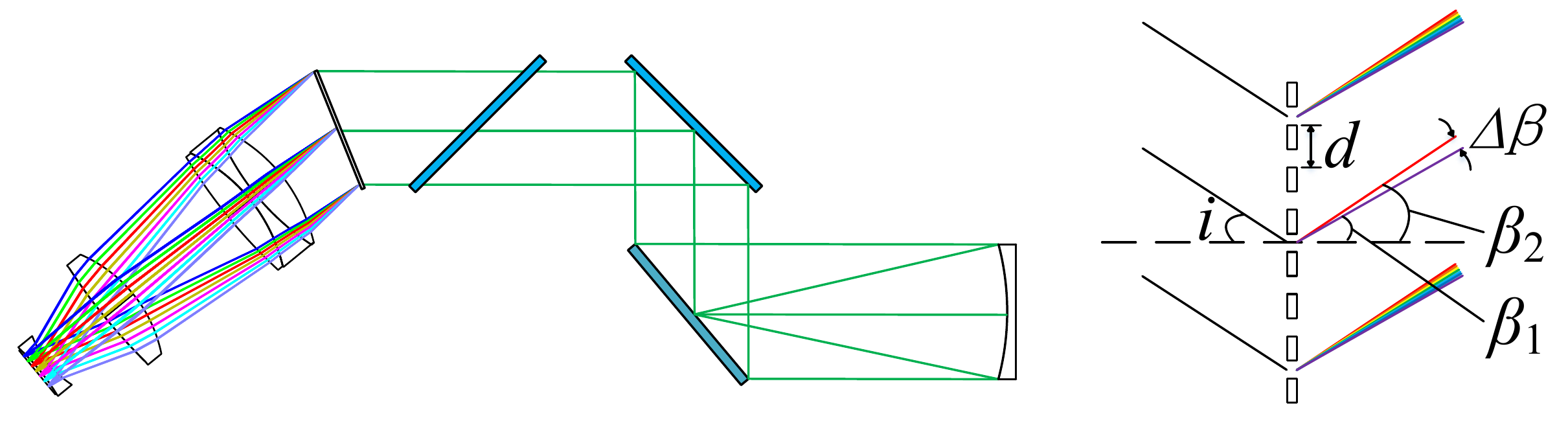}
\caption{(a) Design layout of one channel of MOFFS; (b) Schematic diagram of grating’s dispersion effect.}
\label{fig:5}
\end{figure}

Figure \ref{fig:5}(a) detailedly shows the optical layout of one of the four channels of the spectrometer. VPHG is the kernel optical element of the whole system dispersing the incident light into different angles for the different wavelength of light. 

 Supposing that the diffraction angles $\beta_1$ and $\beta_2$ are for the light with a shorter wavelength $\lambda _1$ and the longer wavelength $\lambda _2$, respectively. Their relation can be specified as

\[\left\{ \begin{array}{l}
d(\sin {\beta _1} + \sin i) = k \cdot {\lambda _1}\\
d(\sin {\beta _2} + \sin i) = k \cdot {\lambda _2}
\end{array} \label{eq:4}\tag{4}\right.,\]
where $d$ is the grating period, $i$ donates the incident angle and $k$ represents the diffraction order. The divergence angle $\bigtriangleup \beta$ of the dispersed beam can be calculated as
\[\Delta \beta  = {\beta _2} - {\beta _1} = \arcsin (\frac{{k \cdot {\lambda _2}}}{d} - \sin i) - \arcsin (\frac{{k \cdot {\lambda _1}}}{d} - \sin i)\label{eq:5}\tag{5}.\]

In this design, only +1 diffraction order is considered, which means that $k =1$ in Eq. (\ref{eq:5}). Since the initial structure of the camera group lens of the spectrometer is rotationally symmetrical about the optical axis, the dispersion angle can be approximated to the field of view of the camera group, which means $\bigtriangleup \beta \approx \theta$. The line density of VPHG for each channel is 
determined based on the conventional choice \cite{Aghamousa:2016sne,BIG}, the incident angle and dispersion angle are also calculated as listed in Table \ref{tab:T4}. These parameters are the basis to determine the initial optical structure of the spectrometer.

\begin{table*}[htbp]
\centering
\caption{\bf Initial parameters of the VPHG.}
\begin{tabular}{ccccc}
\hline
Channel&BLUE&RED&NIR&IR\\
\hline
Wavelength\ range & 360nm-580nm & 560nm-740nm & 720nm-900nm&877nm-1100nm\\
Linear\ density & 900 lines/mm & 1050 lines/mm & 1000 lines/mm&850 lines/mm\\
Incident\ angle & 25.05° & 24.11° & 24.2°&34.67°\\
Grating\ spectral\ dispersion & 11.3632° & 11.2825°& 11.2818°&11.2819°\\
\hline
  \label{tab:T4}
\end{tabular}
\end{table*}

\subsection{Determination of camera group's parameters}
The size of the photosensitive region of 4K×4K CCD equals 61.44mm×61.44mm. Taking the alignment tolerance and edge effects of CCD into consideration \cite{Aghamousa:2016sne}, only the central region of 59mm×59mm is considered. The focal length of the camera group can be obtained as
\[f = a{\left( {2\tan \frac{\theta }{2}} \right)^{ - 1}}\label{eq:6}\tag{6},\]where $a$ is the margin length of the working area of CCD. With the information given above, the value of $f$ is calculated as 298.66mm.

\section{Design and results}
\subsection{Optical layout of the system}
\begin{figure}[ht!]
\includegraphics[width=6in]{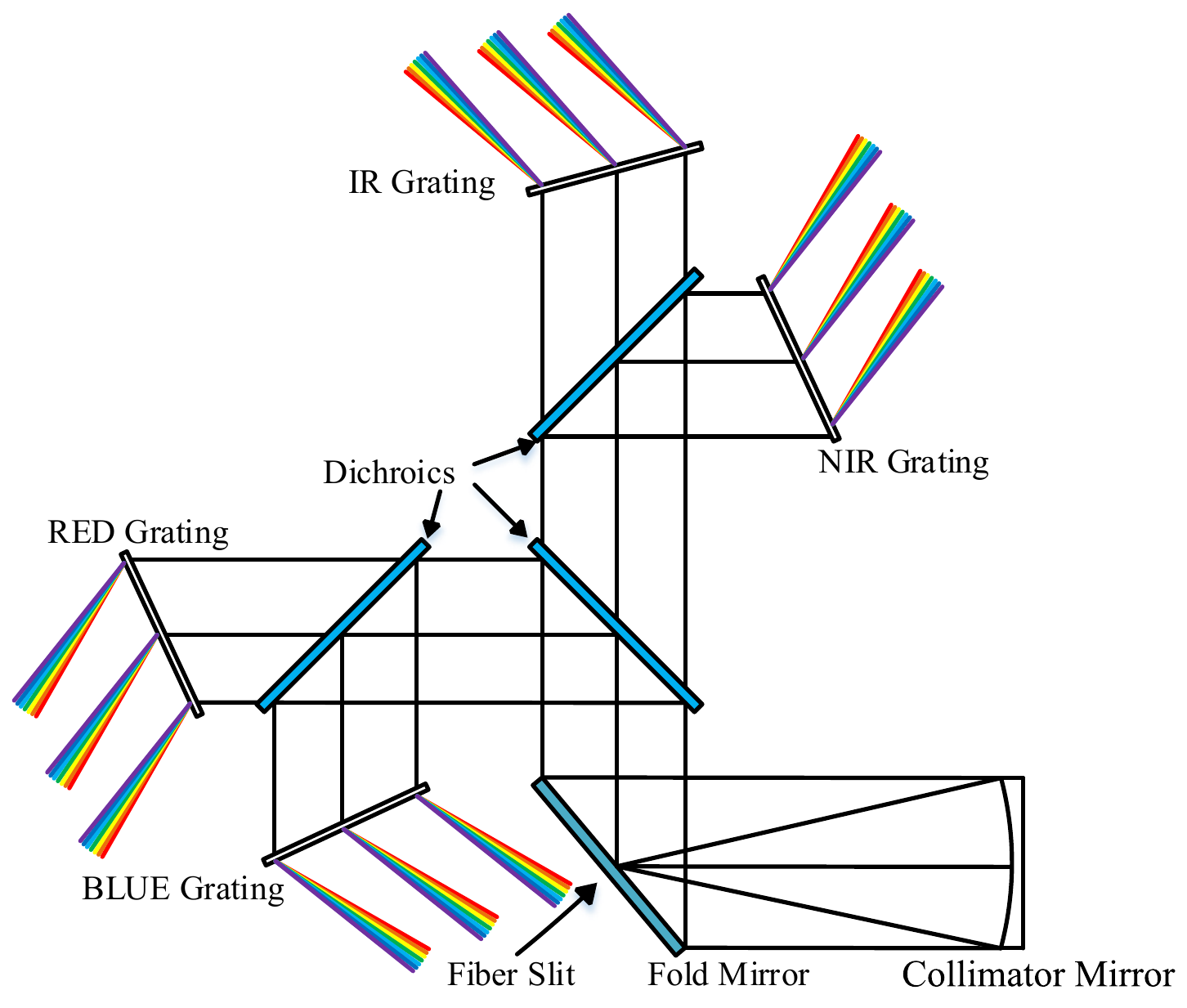}
\caption{Design layout of optical arrangement.}
\label{fig:6}
\end{figure}

As shown in Fig. \ref{fig:6}, the fiber slit is positioned at the optical axis of the spherical collimator mirror. To avoid the vignetting effect caused by the fiber slit as well as the fiber bundle, a flat fold mirror is located physically around the fiber slit. The reflected beam is divided into 4 channels by the 3 dichroics. Since most glasses have a strong spectral absorption around 360nm, the blue channel is designed as a reflective unit before the camera group lens.

\subsection{Design procedures}
The initial optical structure of the camera group optical system is determined by the focal length and the field angle that have been provided in Eq. (\ref{eq:1}) and Eq. (\ref{eq:6}). The optical performance requirements for the camera groups are listed in Table \ref{tab:T5}.

\begin{table*}[htbp]
\centering
\caption{\bf Parameter requirements of the camera group lens.}
\begin{tabular}{ccccc}
\hline
Channel&BLUE&RED&NIR&IR\\
\hline
Wavelength\ range & 360nm-580nm & 560nm-740nm & 720nm-900nm&877nm-1100nm\\
Spectral\ resolution & $\rm <0.057nm/pixel$ & $\rm <0.052nm/pixel$ & $\rm <0.052nm/pixel$&$\rm <0.057nm/pixel$\\
Transmittance& $\rm >86\%(360nm-390nm)$ &$\rm  >90\% $& $\rm >90\%$&$\rm >90\%$\\
\ & $\rm >90\%(390nm-580nm)$&\ &\ &\\\
RMS\ radius &<13$\mu$m&<13$\mu$m&<13$\mu$m&<13$\mu$m\\
Focal\ length& 298.66mm & 298.66mm & 298.66mm&298.66mm\\
\hline
  \label{tab:T5}
\end{tabular}
\end{table*}

\begin{figure}[ht!]
\includegraphics[width=6in]{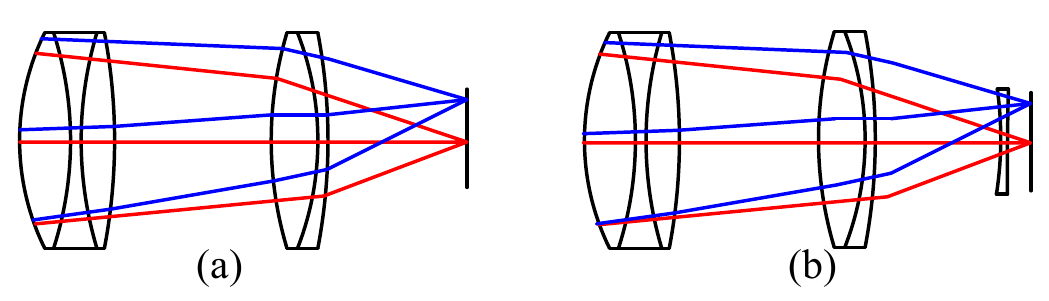}
\caption{Evolution of the initial structure model for the camera group.}
\label{fig:7}
\end{figure}

\begin{figure}[ht!]
\includegraphics[width=6in]{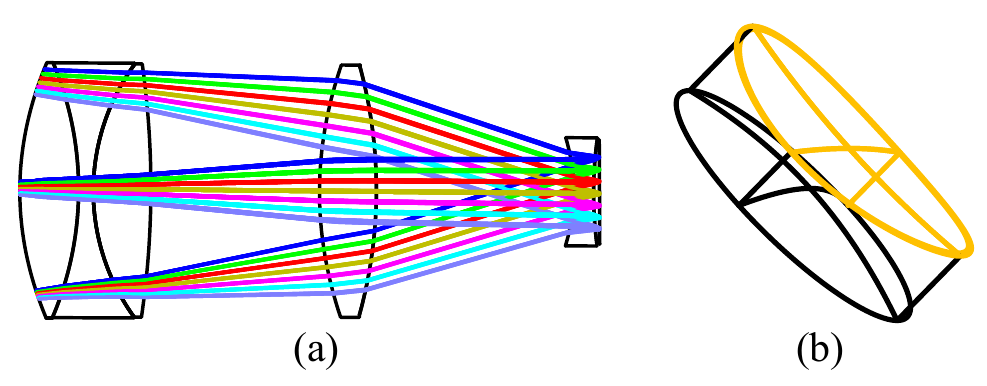}
\caption{Evolution of the initial structure model for the camera group: (a) Camera lens system; (b) Biconical surface of the field flattener.}
\label{fig:8}
\end{figure}

Firstly, we choose the Petzval lens system as the initial structure as shown in Fig. \ref{fig:7}(a), which consists of two positive lenses \cite{1966Objective,1992Modern}. The power of the whole optical system is equally distributed in two lenses. And, a field flattener is usually added near the image plane to improve the optical performance as shown in Fig. \ref{fig:7}(b) \cite{2008Handbook}. To further simplify the design, our design is promoted to have a triplet lens, a single positive lens, and a field flattener as shown in Fig. \ref{fig:8}(a). To correct the field curvature introduced by the collimator mirror, we use a “biconical” surface as the front surface of the field flattener as shown in Fig. \ref{fig:8}(b). The sag height of the biconical surface is expressed as
\[z = \frac{{{c_x}{x^2} + {c_y}{y^2}}}{{1 + \sqrt {1 - \left( {1 + {k_x}} \right)c_x^2{x^2} - \left( {1 + {k_y}} \right){c_y}{y^2}} }},\label{eq:7}\tag{7}\]
where $c_x$ and $c_y$ are curvatures of the surface in $x$ and $y$ directions, and $k_x$ and $k_y$ are the conic coefficients of the surface in $x$ and $y$ directions.
Considering the cost, the spectral absorption characteristics, the refractive index, and the Abbe number, the selected glass combinations for each channel are listed in Table \ref{tab:T6}. Since most glasses have strong absorption for the blue spectrum, the transmittance plays a major role in the glass choice in the BLUE channel. The main reason for choosing fused silica as the material of the last lens is that it is not radioactive \cite{Aghamousa:2016sne}.
\begin{table}[htbp]\small
\centering
\caption{\bf Glass material of camera group lens.}
\begin{tabular}{lllll}
\hline
Channel&BLUE&RED&NIR&IR\\
\hline
Front\ group & H-K9LGT &H-QK3L& H-QK3L&H-QK3L\\
\ &QF8&H-F2& H-F2&H-F2\\
\ &H-QK3L &H-QK3L & H-QK3L&H-QK3L\\
Rear\ group &H-K9LGT &H-K9LGT & H-K9LGT&H-K9LGT \\
Field\ flatter&F\_SILICA&F\_SILICA&F\_SILICA&F\_SILICA\\
\hline
  \label{tab:T6}
\end{tabular}
\end{table}
After the initial structure of the camera group, the VPHG and the collimator mirrors are independently designed, and then we take a further optimization of the whole optical system to improve the optical performance. In the optimization process, we use the diameter of the collimated beam and the size of the spectral imaging area as the constraints. We optimize the line density and the incident angle of the grating while constraining the dispersion angle. The optimization is implemented in ZEMAX \cite{zemax}.

\begin{figure}[ht!]
\includegraphics[width=6in]{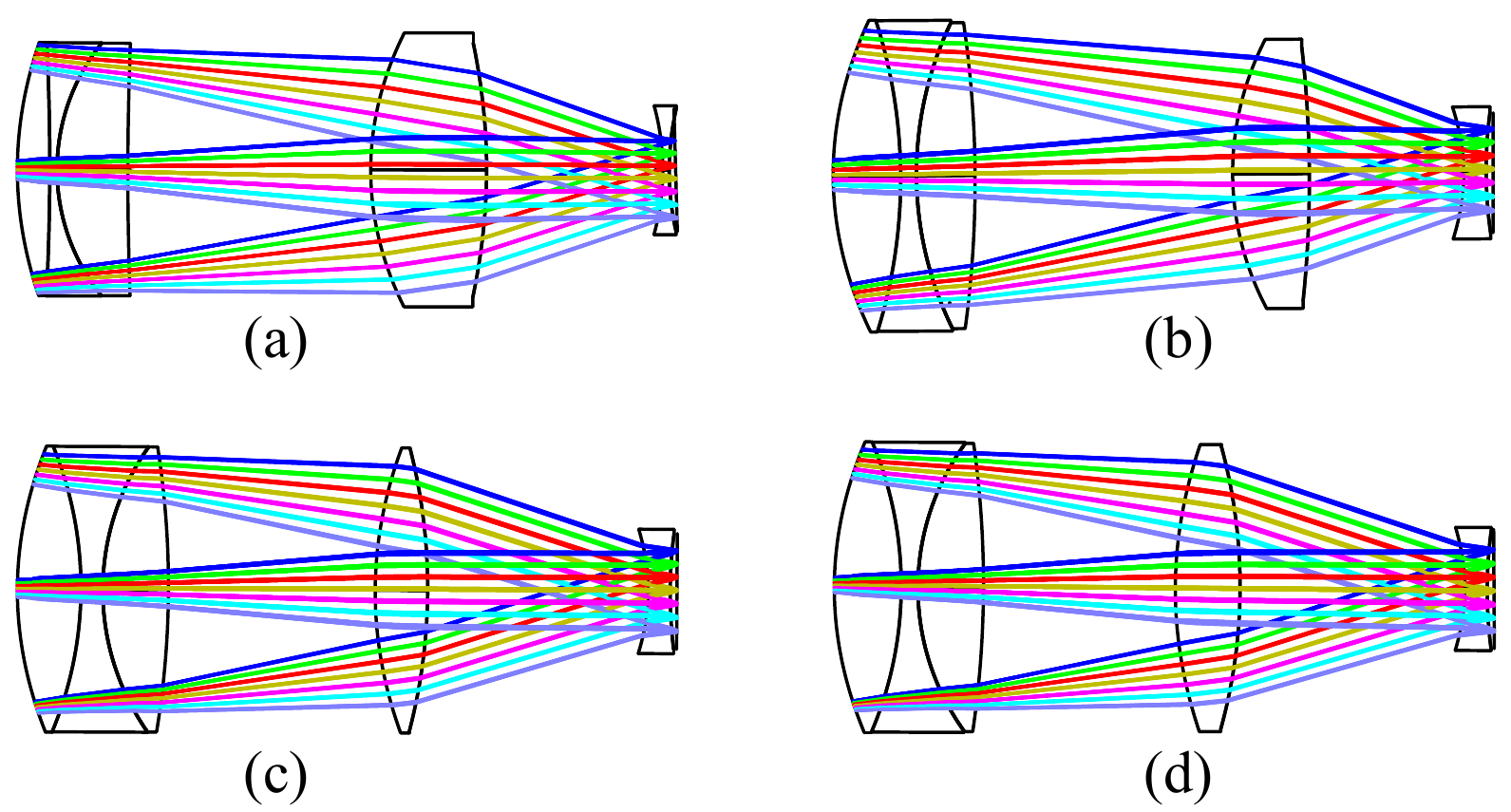}
\caption{2D layout of the optimized camera group lens in each channel:(a) BLUE channel; (b) RED channel; (c) NIR channel; (d) IR channel.}
\label{fig:9}
\end{figure}

\begin{table*}[htbp]
\centering
\caption{\bf Optimized parameters of the VPH gratings.}
\begin{tabular}{ccccc}
\hline
Channel&BLUE&RED&NIR&IR\\
\hline
Wavelength\ range & 360nm-580nm & 560nm-740nm & 720nm-900nm&877nm-1100nm\\
Linear\ density & 901 lines/mm & 1016 lines/mm & 979 lines/mm&792 lines/mm\\
Incident\ angle & 15.11° & 21.432° & 24.452°&24.036°\\
Grating\ spectral\ dispersion & 11.5318° & 10.9880°& 10.9359°&10.9199°\\
\hline
  \label{tab:T7}
\end{tabular}
\end{table*}

\subsection{Design results}
The final obtained camera group lens for each channel are shown in Fig. \ref{fig:9}, and the final grating parameters are listed in Table \ref{tab:T7}. The final optimization parameters of the camera group lens in each channel obtained by the optical design software ZEMAX are shown in Table \ref{tab:T10} to \ref{tab:T13}.

\begin{table}[htbp]
\centering
\caption{\bf Optimization parameters of camera group lens in BLUE channel.}

\begin{tabular}{llllll}
\hline
Element&Material&Curvature&Thickness&Conic\\
 & &Radius(mm)&(mm)& \\
\hline
$\rm Triplet-S1$&H-K9LGT&287.430&25.163&-0.954\\
$\rm Triplet-S2$&QF8&-1957.749&5.643&0\\
$\rm Triplet-S3$&H-QK3L&156.250&54.996&0\\
$\rm Triplet-S4$&—&2588.211&188.565&0\\
$\rm Rear-S1$&H-K9LGT&223.149&90.000&-0.603\\
$\rm Rear-S2^a$&—&-443.488&136.757&-2.641\\
$\rm Flat-S1^c$&F\_SILICA&-191.595&5.000&1\\
$\rm Flat-S2^c$&—&212.592&5.000&0\\
$\rm CCD^d$&—&infinite&—&0\\
\hline
 \label{tab:T10}
 \end{tabular}
 \begin{tabular}{l}
\footnotesize{$^{\rm a}$Aspheric surface coefficients:$\rm \alpha_4= 3.09E-009$,$\rm \alpha_6= -7.72E-014$.}\\
\footnotesize{$^{\rm b}$Biconical surface: $\rm X\ Curvature\ Radius=-182.610$, $\rm X\ Conic=-5$.}\\
\footnotesize{$^{\rm c}$Biconical surface: $\rm X\ Curvature\ Radius= infinite$, $\rm X\ Conic=0$.}\\
\footnotesize{$^{\rm d}$Slit image: $\rm slit\ angle\ with\ X\ axis=1.680^\circ$.}\\
\end{tabular}
\end{table}

\begin{table}[htbp]
\centering
\caption{\bf Optimization parameters of camera group lens in RED channel.}

\begin{tabular}{llllll}
\hline
Element&Material&Curvature&Thickness&Conic\\
 & &Radius(mm)&(mm)& \\
\hline
$\rm Triplet-S1$&H-QK3L&239.989&48.368&-0.843\\
$\rm Triplet-S2$&H-F2&-370.194&11.359&0\\
$\rm Triplet-S3$&H-QK3L&250.850&42.000&0\\
$\rm Triplet-S4$&—&-787.564&184.632&0\\
$\rm Rear-S1$&H-K9LGT&204.668&55.000&-0.192\\
$\rm Rear-S2$&—&-832.669&110.718&0\\
$\rm Flat-S1$&F\_SILICA&-134.887&16.000&-2.786\\
$\rm Flat-S2^a$&&317.771&4.985&0\\
$\rm CCD^b$& &infinite&—&0\\
\hline
 \label{tab:T11}
\end{tabular}
 \begin{tabular}{l}
\footnotesize{$^{\rm a}$Biconical surface: $\rm X\ Curvature\ Radius= 967.739$, $\rm X\ Conic=0$.}\\
\footnotesize{$^{\rm b}$Slit image: $\rm slit\ angle\ with\ X\ axis=-0.334^\circ$.}\\
\end{tabular}
\end{table}

\begin{table}[htbp]
\centering
\caption{\bf Optimization parameters of camera group lens in NIR channel.}
\begin{tabular}{llllll}
\hline
Element&Material&Curvature&Thickness&Conic\\
 & &Radius(mm)&(mm)& \\
\hline
$\rm Triplet-S1$&H-QK3L&260.508&46.493&-1.076\\
$\rm Triplet-S2$&H-F2&-273.735&15.990&0\\
$\rm Triplet-S3$&H-QK3L&180.575&47.999&0\\
$\rm Triplet-S4$&—&-659.152&151.243&0\\
$\rm Rear-S1$&H-K9LGT&287.219&38.352&-0.886\\
$\rm Rear-S2$&—&-413.081&161.364&0\\
$\rm Flat-S1$&F\_SILICA&-138.478&16.001&-2.586\\
$\rm Flat-S2^a$&&335.936&4.971&0\\
$\rm CCD^b$& &infinite&—&0\\
\hline
 \label{tab:T12}
\end{tabular}
 \begin{tabular}{l}
\footnotesize{$^{\rm a}$Biconical surface: $\rm X\ Curvature\ Radius= 603.746$, $\rm X\ Conic=0$.}\\
\footnotesize{$^{\rm b}$Slit image: $\rm slit\ angle\ with\ X\ axis=-0.148^\circ$.}\\
\end{tabular}
\end{table}

\begin{table}[htbp]
\centering
\caption{\bf Optimization parameters of camera group lens in IR channel.}
\begin{tabular}{llllll}
\hline
Element&Material&Curvature&Thickness&Conic\\
 & &Radius(mm)&(mm)& \\
\hline
$\rm Triplet-S1$&H-QK3L&249.801&48.005&-0.979\\
$\rm Triplet-S2$&H-F2&-267.096&12.000&0\\
$\rm Triplet-S3$&H-QK3L&175.583&47.006&0\\
$\rm Triplet-S4$&—&-727.455&137.670&0\\
$\rm Rear-S1$&H-K9LGT&299.867&46.438&-1.304\\
$\rm Rear-S2$&—&-383.637&161.330&0\\
$\rm Flat-S1$&F\_SILICA&-135.802&16.000&-2.914\\
$\rm Flat-S2^a$&&320.043&4.845&0\\
$\rm CCD^b$& &infinite&—&0\\
\hline
 \label{tab:T13}
\end{tabular}
 \begin{tabular}{l}
\footnotesize{$^{\rm a}$Biconical surface: $\rm X\ Curvature\ Radius= -598.867$, $\rm X\ Conic=0$.}\\
\footnotesize{$^{\rm b}$Slit image: $\rm slit\ angle\ with\ X\ axis=0.474^\circ$.}\\
\end{tabular}
\end{table}

\subsection{Optical performance}
\begin{figure}[ht!]
\includegraphics[width=6in]{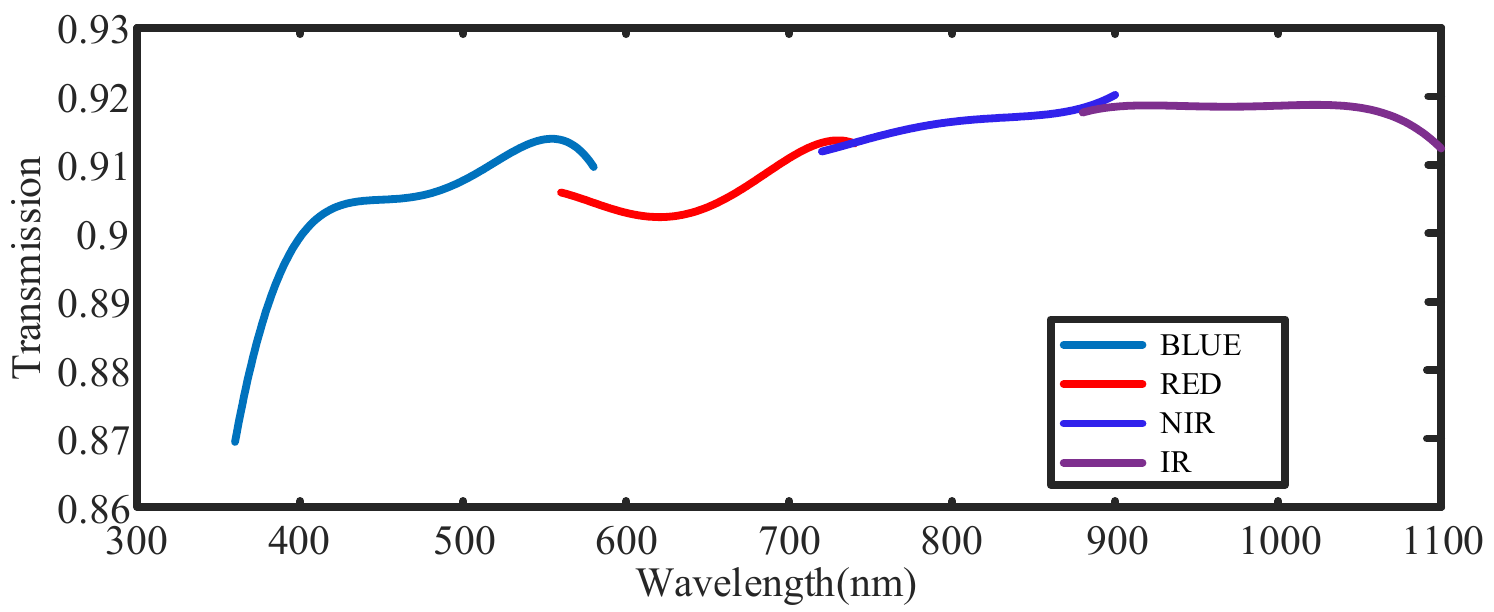}
\caption{The throughput of the camera group lenses. The blue, red, green and purple curves are for the BLUE, RED, NIR and IR channels respectively.}
\label{fig:10}
\end{figure}

The transmittance is one of the most important metrics of optical performance for the spectrograph system. Since the system has no vignetting effect, we mainly consider two loss factors for the system, i.e. the absorption of glasses and the reflection loss on the optical surfaces. Figure \ref{fig:10} shows the transmittance curves of the four channels. The anti-reflection coatings on the lenses are assumed to have a transmission of 99\% in all channels. The result shows that the transmittance meets the requirements given in Table \ref{tab:T5}. 

\begin{figure*}[ht!]
\includegraphics[width=6in]{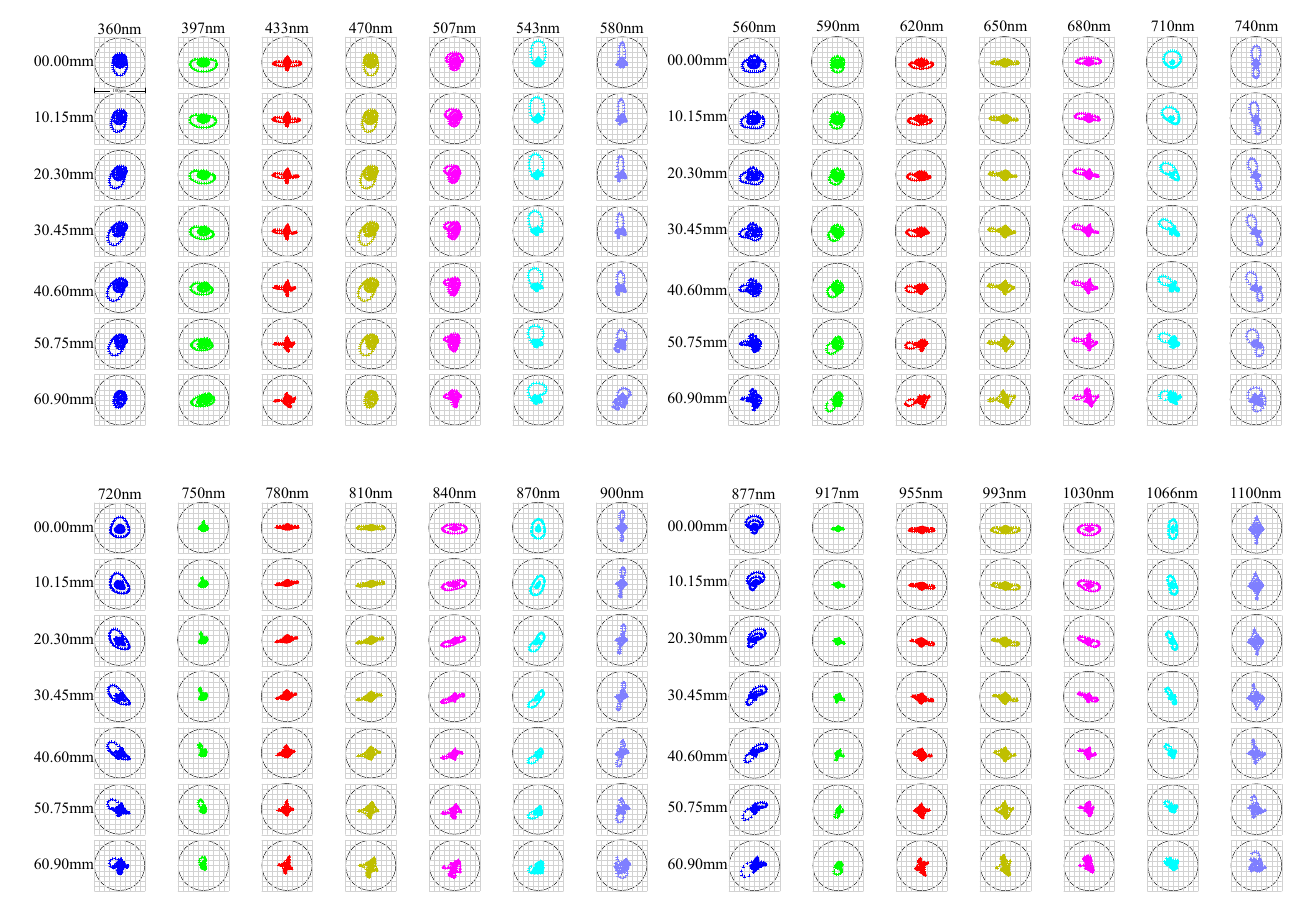}
\caption{The ray traced spot diagram for the spectrograph.}
\label{fig:11}
\end{figure*}

\begin{figure}[ht!]
\includegraphics[width=6in]{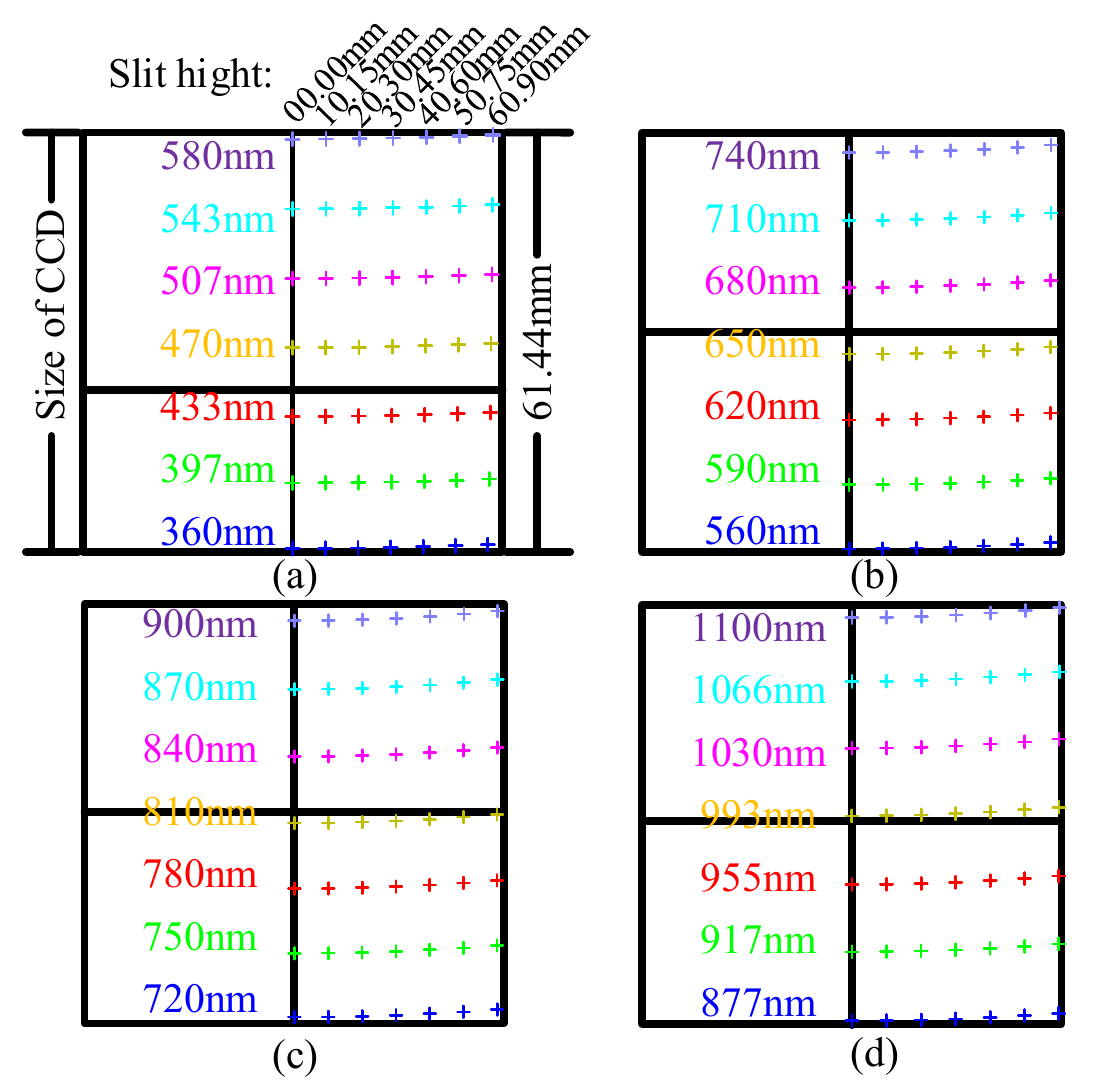}
\caption{The 2D layout of focal plane of each channel.}
\label{fig:12}
\end{figure}

As shown in Fig. \ref{fig:11}, the RMS radius is less than 12.2 $\mu$m for the RED, NIR, and IR channels, and is less than 13 $\mu$m for the BLUE channel. The 2D layout of the focal plane is shown in Fig. \ref{fig:12}. We can find that the distortions of the spectral images are well controlled(the final optimization parameters of each channel obtained by the optical design software ZEMAX are provided in the appendix). And these results meet the predefined requirements.

\section{Tolerance Analysis}
To make the spectrometer be friendly to fabrication and alignment, the tolerance budgets are eased in ZEMAX referring to Table \ref{tab:T8}. A thousand trials of Monte Carlo tolerancing have been performed, and Table \ref{tab:T9} lists the five cases which have the biggest negative impact on RMS radius of each channel. Analysis results show that the RMS radius of each channel with 90\% confidence is less than 15 $\mu$m for our proposed error budgets.

\begin{table}[htbp]\small
\centering
\caption{\bf Tolerances setting in ZEMAX of the calculation.}
\begin{tabular}{llllll}
\hline
Tolerances&BLUE&RED&NIR&IR\\
\hline
Surface\ tolerances&\ &\ &\ &\ \\
Radius&±0.05mm&±0.05mm&±0.05mm&±0.05mm\\
Thickness&±0.05mm&±0.05mm&±0.05mm&±0.05mm\\
Decenter\ in\ X&±0.05mm&±0.05mm&±0.05mm&±0.05mm\\
Decenter\ in\ Y&±0.05mm&±0.05mm&±0.05mm&±0.05mm\\
Tilt\ in\ X&±0.01°&±0.05°&±0.03°&±0.05°\\
Tilt\ in\ Y&±0.01°&±0.05°&±0.03°&±0.05°\\
Element\ tolerances&\ &\ &\ &\ \\
Decenter\ in\ X&±0.05mm&±0.05mm&±0.05mm&±0.05mm\\
Decenter\ in\ Y&±0.05mm&±0.05mm&±0.05mm&±0.05mm\\
Tilt\ in\ X&±0.01°&±0.05°&±0.03°&±0.05°\\
Tilt\ in\ Y&±0.01°&±0.05°&±0.03°&±0.05°\\
Index\ tolerances&\ &\ &\ &\ \\
Index&±0.0003&±0.0005&±0.0004&±0.0005\\
Abbe&1\%&1\%&1\%&1\%\\
\hline
  \label{tab:T8}
\end{tabular}
\end{table}

\begin{table}[htbp]\footnotesize
\caption{\bf Tolerances setting in ZEMAX of the calculation.}
\begin{tabular}{lllll}
\hline
Offenders&$\rm Increase\ of$\\
&$\rm RMS\ radius(\mu m)$\\
\hline
BLUE Channel&\ &\ &\ &\ \\
$\rm TIND^a\ after\ Triplet-S2$&1.5956\\
$\rm TIND\ after\ Triplet-S3$&1.4958\\
$\rm TIND\ of\ the\ rear\ lens$&0.9982\\
$\rm TETY^b\ of\ the\ Fold\ Mirror$&0.5327\\
$\rm TETY\ of\ the\ NIR\ and\ IR\ Pass\ Dichroic$&0.5170\\
RED Channel&\ &\ &\ &\ \\
$\rm TETY\ of\ the\ Triplet\ lens$&2.7588\\
$\rm TIND\ after\ Triplet-S2$&2.2484\\
$\rm TETY\ of\ the\ Rear\ lens$&2.1757\\
$\rm TETX^c\ of\ the\ Rear\ lens$&1.8139\\
$\rm TIND\ after\ Triplet-S1$&1.7205\\
NIR Channel&\ &\ &\ &\ \\	
$\rm TIND\ after\ Triplet-S2$&3.4971\\
$\rm TETX\ of\ the\ Rear\ lens$&2.3165\\
$\rm TIND\ after\ Triplet-S3$&1.8773\\
$\rm TETY\ of\ the\ Rear\ lens$&1.6681\\
$\rm TIND\ of\ the\ Rear\ lens$&1.4419\\
IR Channel&\ &\ &\ &\ \\	
$\rm TIND\ after\ Triplet-S2$&4.1641\\
$\rm TETY\ of\ the\ Rear\ lens$&3.3088\\
$\rm TETX\ of\ the\ Rear\ lens$&2.9817\\
$\rm TIND\ after\ Triplet-S3$&2.0441\\
$\rm TETY\ of\ the\ Fold\ Mirror$&1.7186\\
\hline
  \label{tab:T9}
\end{tabular}
 \begin{tabular}{l}
\footnotesize{$^{\rm a}$TIND:tolerance on index or refraction.}\\
\footnotesize{$^{\rm b}$TETY:tolerance on element tilt about the y axis.}\\
\footnotesize{$^{\rm c}$TETX:tolerance on element tilt about the x axis.}\\
\end{tabular}
\end{table}

\section{Conclusion}
To summarize, we propose a MOFFS system for the wide-field spectroscopic survey telescopes, especially for China’s SSST. The proposed design is demonstrated to have superior optical performance with good image quality, high optical throughput, and low cost of glass materials. A brief tolerance analysis of the proposed design indicates a moderate error budget for optical alignment and optical surface quality, which demonstrates its good manufacturability. And we believe that the design ideas, including the calculation of initial structures, optimization procedures of the VPHG parameters as well as the camera groups, can provide a good reference and guidance to researchers for the future instrumentation of such kind of astronomical spectrograph systems.

\medskip

\noindent\textbf{Funding.} National Natural Science Foundation of China (61805088); Science, Technology, and Innovation Commission of Shenzhen Municipality (JCYJ20190809100811375); Key Research and Development Program of Hubei Province (2020BAB121); Fundamental Research Funds for the Central Universities (2019kfyXKJC040); Innovation Fund of WNLO.

\medskip

\noindent\textbf{Disclosures.} The authors declare no conflicts of interest.




\bibliography{sample}



\bibliographystyle{unsrt}  

\end{document}